\documentclass{article}

\usepackage[final,nonatbib]{neurips_2022}

\usepackage[utf8]{inputenc} 
\usepackage[T1]{fontenc}    
\usepackage{hyperref}       
\usepackage{url}            
\usepackage{booktabs}       
\usepackage{amsfonts}       
\usepackage{nicefrac}       
\usepackage{microtype}      
\usepackage{xcolor}         
\usepackage{enumitem}
\usepackage{comment}

\newcommand{\vinod}[1]{{\color{red}#1 - VP}}
\newcommand{\rida}[1]{{\color{orange}#1 - RQ}}

\title{Cultural Incongruencies in Artificial Intelligence}

\author{
  Vinodkumar Prabhakaran\\
  Google Research\\
  San Francisco, US \\
  \texttt{vinodkpg@google.com} \\
   \And
   Rida Qadri \\
   Google Research\\
  San Francisco, US \\
   \texttt{ridaqadri@google.com} \\
   \And
   Ben Hutchinson \\
   Google Research\\
   Sydney, Australia \\
   \texttt{benhutch@google.com} \\
}

\begin{document}
\maketitle

\begin{abstract}
Artificial intelligence (AI) systems attempt to imitate human behavior. How well they do this imitation is often used to assess their utility and to attribute human-like (or artificial) intelligence to them. However, most work on AI refers to and relies on human intelligence without accounting for the fact that human behavior is inherently shaped by the cultural contexts they are embedded in, the values and beliefs they hold, and the social practices they follow. Additionally, since AI technologies are mostly conceived and developed in just a handful of countries, they embed the cultural values and practices of these countries. Similarly, the data that is used to train the models also fails to equitably represent global cultural diversity. Problems therefore arise when these technologies interact with globally diverse societies and cultures, with different values and interpretive practices. In this position paper, we describe a set of cultural dependencies and incongruencies in the context of AI-based language and vision technologies, and reflect on the possibilities of and potential strategies towards addressing these incongruencies.
\end{abstract}

\section{Introduction}
\label{sec_intro}
\label{sec_culture_and_ai}

Artificial Intelligence (AI) research includes the demonstration by machines of human-like intelligence and capabilities, measured in terms of how well they match human behavior on certain tasks. 
This is reflected in current AI frontiers including language understanding and generation, image understanding and creation, knowledge representation and reasoning, and the automatic acquisition (Machine Learning) of capabilities that support these and other domains.
However, human behavior is inherently shaped by the cultural contexts humans are embedded in, the values and beliefs they hold, and the social practices they follow. Thus, any AI system mimicking human behaviour will reflect this culturally-shaped understandings of behaviour. 
However problems can arise when the culture that develops and shapes the AI differs from the globally diverse cultures of the human-AI interaction contexts, due to tensions and misalignments between cultural values and practices. 
We posit the following two foundational questions:
(1) Which aspects of AI systems are dependent on culture, for instance, an AI system's appropriateness for a given ecosystem may depend substantially on the cultures of the humans in that ecosystem?, and (2) What incongruencies and harms emerge when there is a mismatch between an AI system's implicit cultural predispositions and the cultural ecosystem it is used within?
In this paper we expand on these questions, and examine the limits of and strategies towards building culturally cognizant AI systems.

\section{Cultural incongruencies and associated harms}

Cultural dependencies of AI systems are rarely accounted for in current AI research and development work.
It is beyond the scope of this short position paper to summarize the myriad existing definitions, across many disciplines, of the term  ``culture'' (for an overview, see, e.g., \cite{rapport2014social,metcalf2006anthropology}). 
First, we focus on cultures created within broader societies demarcated geographically through national and regional boundaries and not cultures of, e.g., specific organizations. Secondly, we focus on those aspects of culture that exhibit significant variation across human societies, including worldviews, belief systems, and social practices. Due to the central importance of communication and interpretive practices within culture \cite{hall1959silent, geertz1973interpretation}, it follows immediately
that communicative and interpretive AI technologies, such as NLP and computer vision, have deep cultural dependencies at various levels.

At a high-level, we can distinguish two ways in which culture interacts with AI systems: \textit{in development} and \textit{in use}. The development process of AI systems interface with culture both through the data and resources that capture culturally shaped human behavior, as well as through the cultural norms and values embodied by the developers and researchers themselves. For instance, modern AI systems that are trained or pre-trained on web data may capture various modalities of human behavior, including language use and images, which implicitly bakes in various cultural aspects that then influence downstream applications. Since language and symbols, and ontology and axiology, play a critical role in the development of AI systems---e.g., through ``labels'' on data, and how ``knowledge'', ``objectivity'' \cite{daston2021objectivity}, ``reality/truth'' \cite{searle1995construction}, and ``system objectives'' are constructed---the cultural norms of the AI developers and researchers also pervasively infuse the AI systems \cite{forsythe2001studying, denton2021genealogy,sambasivan2021re}. 

On the other hand, how AI systems are used, whether they perform the tasks they are built for in ways that adhere to culturally shaped expectations, and how they interact with other human behaviours are all culturally dependent. For instance, interpretive tasks are inherently shaped by the culture within which they are embedded in, including not only the cultural-linguistic dependencies \cite{hershcovich2022challenges,hovy2021importance} of tasks such as inferring emotion, sentiment, offensiveness, but also image and symbol interpretation---including gestures, facial expressions, taboo imagery including pornography and violence, and denotations and connotations of symbols \cite{ekman1973cross,chan2007consumers,cohn2013visual,chandler2007semiotics}.
When the cultural assumptions and norms that are baked into the AI systems during its development are at odds with the cultural norms and expectations of the target cultural ecosystems, we see breakdowns and failures such as cultural misinterpretations or cultural misrepresentations, which we collectively call \textit{cultural incongruencies}. 
In this section, we present five kinds of harms cultural incongruencies may cause:

\begin{itemize}[itemsep=2pt,topsep=0pt,leftmargin=*]

    \item \textbf {Cultural barriers}: Not accounting for cultural biases in training data often result in disparate performance of AI systems across different cultural contexts, often disadvantaging cultures that are already historically marginalized. For instance, failing to understand or generate certain languages and dialects may cause NLP-based virtual assistants to perform poorly for users who use those languages or dialects. Similarly, question-answering systems may perform worse on questions related to cultural artifacts from certain cultures, owing both to disparities in training data as well as to gaps in any underlying ontologies and databases. Another example is a computer vision system failing to detect or generate objects, events, or movements that are specific to certain cultures; e.g. a woomera (Australian spear thrower) in a photo, or a description-to-depiction text-to-image system rendering better quality images of cultural artifacts specific to one culture than another. 
    
    \item \textbf{Imposing hegemonic classifications}: The cultural categories of AI developers can become embedded in AI systems and then applied to diverse cultural contexts, imposing epistemic practices that are not endemic to the local cultural context \cite{foucault1989order}.
    Such categorizations using the classification schemes of the developers' culture can silence or minimize local cultural perspectives while valorizing the hegemonic culture \cite{bowker2000sorting}.

    \item \textbf{Safety gaps}: With increasing adoption of AI systems, there are also increasing efforts on ensuring the AI systems are safe and fair \cite{dinan2021anticipating,thoppilan2022lamda}. However, these safety guardrails fail if they don't account for the target cultural ecosystems \cite{sambasivan2021re}. For instance, content moderation systems meant to detect offensiveness and misinformation may miss culture-specific offensive terms and interpretations allowing toxic or violent speech to propagate for some cultural settings \cite{prabhakaran2020online,ghosh2021detecting}. Pedestrian detection systems trained and tested on Western streets may not be effective in cities in the Global South as rules of mobility e.g. what it means to honk and where is it acceptable to cross a road are created collectively within cultures and differ significantly around the world. 

    \item \textbf{Violating cultural values}: Lack of consideration for the cultural context in which an AI system is to be deployed may result in violating the norms that are important to those communities \cite{johnson2022ghost}. For instance, a generative language model may produce text that are offensive within certain cultures, even if the language is deemed appropriate at large, e.g., mixing words that are sacred with words that are considered profane. Similarly, a computer vision system may violate cultural norms by producing labels or captions that differ from those preferred by members of that culture.
    
    \item \textbf {Cultural erasure}: Cultural erasure occurs when knowledge, histories, and identities of a particular people are erased either through omission, trivialization or simplification \cite{tuchman1976mass}. \cite{Gerbner1970} describes such erasure as `symbolic annihilation'; i.e., by not being represented, cultures are annihilated from memory if not physically then metaphorically. Such erasure can happen when technologies homogenize diversity of cultural lives,  creating simplified caricatures e.g. a text-to-image model rendering a mosque when prompted to symbolize Islam, not recognizing that Islam is a political, historic, artistic or geographical term not just a religious one. Erasure harm is especially problematic in the context of pre-trained models where such erasure is then also propagated to downstream models.

 \end{itemize}

\section{A Research Agenda Towards Culturally Cognizant AI}

Having outlined the various incongruencies and associated harms that might arise when AI systems interface with cultural contexts, we now turn to a set of high-level concrete research directions that can begin to mitigate these cultural incongruencies. A perfect culturally competent AI system is not the goal here, as culture is not a static variable that can be easily encoded into a technology, rather a complex and dynamic system that is constantly being transmitted and transformed. Instead, we lay out research questions as openings/opportunities to start a conversation on what a culturally cognizant AI system might look like, is it a desirable goal, and if so, how we may get closer to that state.

\vspace{-7pt}
\paragraph{How do we identify and measure cultural harms of AI systems:} The domain of `culture' 
has long been studied by anthropologists and sociologists.
Yet, as we have argued in this paper cultural dependencies and incongruencies of AI technology call for a focus of AI/ML researchers on measuring cultural harms of emerging technologies. This task would require: developing conceptualizations and operationalizations of culture without falling into the trap of positivism; creating theoretically rigorous and community driven metrics for measuring complex phenomena like cultural erasure; mapping the incongruencies like the ones we highlighted to on-ground harms for cultural systems. This research question can not be undertaken by AI researchers alone, but requires interdisciplinary collaborations with scholars who have long studied the interplay between culture and technology as well as centering the experiences of communities whose cultural systems we seek to study.

\vspace{-7pt}
\paragraph{How can we build culturally situated evaluations:}
In order to ensure that AI system evaluations are culturally situated, it is important to evaluate if the test data reflects the distribution or interpretive practices of the ecosystems in which it is used \cite{hutchinson2022evaluation}. Fairness evaluations framed in the West may not readily apply to other socio-cultural contexts \cite{sambasivan2021re}. Resources and adversarial probes used for testing may not have cross-cultural coverage; for instance, identity terms (e.g., \textit{African American}) used to test for racial biases in language models are often framed within the US context, but they do not capture the axes of discrimination across the globe. Cultures may also differ in the relative importance they place on different evaluation metrics (e.g., the relative weightings of false positives and false negatives), or of how much they value average-case vs. worst-case behavior. More research is needed to build evaluation paradigms that can effectively incorporate such multi-cultural considerations.

\vspace{-7pt}
 \paragraph{How can diverse perspectives be integrated into the AI pipeline:} Many incongruencies we discussed emerge from the narrowness of perspectives represented in the training data and labels, and in the values underlying AI development.  To what extent can we diversify these perspectives and what impact would these mitigation strategies have? For instance, human raters form a major part of AI workforce, but they are often deemed interchangeable without accounting for the diverse socio-cultural perspectives they bring to the rating task(s) \cite{diaz2022crowdworksheets}.  
 Replicating data labeling across the all socio-cultural backgrounds across the globe poses challenges at scale; more research is needed to effectively and efficiently incorporate diverse cultural perspectives in rater pools.
 Similarly, the AI research and development workforce is also not representative of the global cultures. Technologies emerge from the particular norms and cultures of these developers and institutions.
 Envisioned AI solutions are influenced by their decisions, such as problem formulation, categorization schemes, and objective functions, to name a few. To what extent, can the AI development processes be opened up to more participatory processes that elevate the voices of more communities as co-creators?

Culture is in of itself a complex phenomena that may not necessarily lend itself to being encoded within technologies, as we imagine them today. Similar conversations have occurred on the possibility of machines encoding fluid social concepts such as identity or gender, and debates exist if these can be understood by machines that are geared towards simplification and objectification. Our call for a research agenda for culturally-cognizant AI is thus aimed towards better understanding the possibilities and limits of encoding cultural cognizance in our AI systems. 

\section*{Acknowledgements}

We would like to thank Courtney Heldreth, Chelsey Fleming, Andrew Smart, and Madeleine Clare Elish for the insightful discussions on this topic which has helped shape this paper.

{\small
\bibliographystyle{plain}
\bibliography{neurips_2022}
}

\end{document}